\documentclass[fleqn,twoside]{article}
\usepackage[headings]{espcrc2}

\readRCS
$Id: espcrc2.tex,v 1.2 2004/02/24 11:22:11 spepping Exp $
\usepackage{graphicx}
\usepackage{amsmath,amssymb}
\usepackage{cite}
\usepackage{srcltx}
\usepackage{axodraw,dsfont}
\usepackage[figuresright]{rotating}

\renewcommand{\d}{\mathrm{d}}

\newcommand{\be}{\begin{equation}}
\newcommand{\eps}{\epsilon}
\newcommand{\dps}{\displaystyle}
\newcommand{\ee}{\end{equation}}
\newcommand{\bea}{\begin{eqnarray}}
\newcommand{\eea}{\end{eqnarray}}
\newcommand{\loopint}[1]{\int \!\!\! \frac{d^D #1}{\left(2\pi\right)^D}\!}
\newcommand{\ESGamma}{S_{\Gamma}}

\newcommand{\rk}{\right)}
\newcommand{\lek}{\left[}
\newcommand{\rek}{\right]}
\newcommand{\hs}[1]{\hspace*{#1 pt}}
\newcommand{\vs}[1]{\vspace*{#1 pt}}

\newcommand{\MeijerG}[7]{G^{#1}_{#2} \bigg( #3 \left|
\begin{array}{c}
\left\{ #4 \right\} \, , \, \left\{ #5 \right\} \\
\left\{ #6 \right\} \, , \, \left\{ #7 \right\}
\end{array}
\rk}
\newcommand{\MB}[2]{\hs{-12} \int\limits_{\hs{15}_{ #1 -i \,
\infty}}^{\hs{15}^{ #1 +i\, \infty}} \hs{-15} \frac{d #2}{2\pi i}}
\newcommand{\pFq}[5]{\, \! _{#1} F_{#2}( #3 \, ; \, #4 \, ; \, #5 )}
\newcommand{\nnb}{\nonumber}
\title{{\vspace*{-10mm}\hfill\footnotesize PITHA-08/16 \\ \vspace*{-2mm} \hfill\footnotesize SFB/CPP-08-41}\\ Hypergeometric functions with rational arguments}

\author{T.~Huber\address[MCSD]{Institut f\"ur Theoretische Physik E, RWTH Aachen University, \\ 
        D-52056 Aachen, Germany}%
        \thanks{Work supported by Deutsche Forschungsgemeinschaft, SFB/TR~9.}}
       
\runtitle{Hypergeometric functions with rational arguments}
\runauthor{T.~Huber}

\begin{document}

\begin{abstract}
We elaborate on the expansion of hypergeometric functions ${}_PF_{P-1}$ about rational parameters, where we focus mainly on the
integer and half-integer case. The strategy and the basic steps of a recently developed algorithm for the expansion about
half-integer parameters are described. The algorithm is implemented in the Mathematica package {\tt HypExp}, by means of which we
derive (partially new) results of selected multi-loop Feynman diagrams. Moreover, we give a new formulation of a conjecture in the
context of a three-loop master integral in HQET.
\vspace{1pc}
\end{abstract}

\maketitle
\section{INTRODUCTION}
Hypergeometric functions (HFs) have a long history in science. Their trail of success already started in the 17th and 18th century when
former days' scientists like Wallis, Leibniz, Newton, Stirling, and Euler discovered hypergeometric-like patterns in sequences of
numbers, mostly in the context of geometrical problems. In 1797, Pfaff and Gau{\ss} observed that the second order differential equation
\bea
x \, (1-x) \, y^{\prime\prime}(x) +\left[c-(a+b+1) \, x\right] \, y^{\prime}(x)  && \nnb \\
- a \, b\, y(x)&=&0 \nnb\\
\eea
is solved by
\be
y(x)= A \; {}_2F_1 \left(a,b;c;x\right)\nnb
\ee
\be
\hs{5}+ B \; x^{1-c} \, {}_2F_1 \left(1+a-c,1+b-c;2-c;x\right) ,
\ee
where
\be
{}_2F_{1}(a,b;c;x)=\sum\limits_{i=0}^\infty\frac{\Gamma(a+i)\Gamma(b+i)\Gamma(c)}{\Gamma(a)\Gamma(b)\Gamma(c+i)}\frac{x^i}{i!}
\ee
represents the hypergeometric series. Later on in the 19th century, argument transformations for the ${}_2F_{1}$-function were derived
(Kummer), and generalizations of the series to other ${}_pF_{q}$ and to two arguments (Appell, Schwarz,
Riemann, Kamp\'{e} de F\'{e}riet) were developed. Nowadays, HFs have a wide range of applications in physics, mathematics, engeneering,
and economics. In the field of particle physics HFs appear in loop and phase space integrals in the context of dimensional
regularization, where usually the regularization parameter $\eps$ appears in the parameters, whereas masses and kinematic
invariants constitute the argument of the HF. Due to the need of extracting poles, finite parts and higher orders in the $\eps$-expansion in
order to ultimately make predictions for physical observables, one is oftentimes confronted with the task of expanding HFs about their
parameters.

Systematic approaches to the expansion of HFs about integer-valued parameters have been developed
\cite{Moch:2001zr,Weinzierl:2004bn,Kalmykov:2007pf} and have been implemented in GiNaC~\cite{Weinzierl:2002hv,Bauer:2000cp},
Mathematica~\cite{Huber:2005yg}, and FORM~\cite{Moch:2005uc}.

In computations involving massive particles~%
\cite{Davydychev:1992mt,Broadhurst:1993mw,Tarasov:2006nk,Davydychev:1998si,Fleischer:2003rm,Jegerlehner:2002em,Jegerlehner:2003py,Davydychev:2000na,Davydychev:2003mv,Schroder:2005va,Bejdakic:2006vg,Grozin:2007ap,Argeri:2007up} the
HFs can contain half-integer parameters. Methods have been developed to expand HFs with
half-integer parameters~\cite{Weinzierl:2004bn,Kalmykov:2006pu,Kalmykov:2006hu,Kalmykov:2007dk}. Recently, we implemented a new
algorithm for the expansion about half-integer parameters in the existing Mathematica package {\tt
HypExp}~\cite{Huber:2005yg,Huber:2007dx}. In the following we briefly describe this algorithm and give examples and applications of its
usage. We conclude with a brief summary on the expansion of HFs about other rational parameters.
\section{ALGORITHM}
We start the explanation of our algorithm with a definition. A HF ${}_PF_{P-1}(\{A_i\};\{B_j\};x)$ is said to be of \textit{type}
$P^r_s$ if, at $\epsilon=0$, $r$ out of the $A_i$ and $s$ out of the $B_j$ are half-integers and all the others are integers.
The algorithm, which was presented for the first time in Ref.~\cite{Huber:2007dx} consists of three parts. \textit{i)~Reduction}.~We express
a HF of a given type in terms of integration and differentiation operators acting on one specific HF of the same type, the latter we
call the basis function of this type, see Eq.~(\ref{eq:final211}). \textit{ii)~Expansion of the basis function in $\eps$}.
The choice of the basis function for each type is not unique and we choose it such that its expansion in $\eps$ can be performed as
conveniently as possible. \textit{iii)~Application of the integration and differentiation operators}. In the last step we have to find
integration and differentiation routines which act on the \textit{expanded} basis function.

Below, we cover each of these steps in
turn. Other, related algorithms which are valid for the integer and/or half-integer case rely on 
the reduction to a set of basis functions by means of recurrence
relations~\cite{Huber:2005yg,Kalmykov:2006pu,Kalmykov:2006hu,Kalmykov:2007dk} or on 
the nested (harmonic and binomial) sums approach~\cite{GonzalezArroyo:1979df,Vermaseren:1998uu,Blumlein:2003gb,Moch:2001zr,Weinzierl:2004bn}.
\subsection{Reduction}
We start the reduction part of the algorithm by introducting a few more definitions. 
We define the short-hand notation
\bea\label{eq:a:bproducts}
\textstyle\prod\limits_{j}^{a:a}=1,\quad\prod\limits_{j}^{a:b}f(j)&=&\textstyle\prod\limits_{j=a}^{b-1}f(j) \quad\textnormal{if}\quad a<b,\nnb\\
\textstyle\prod\limits_{j}^{a:b}f(j)&=&\textstyle\prod\limits_{j=b}^{a-1}\displaystyle\frac{1}{f(j)} \quad\textnormal{if}\quad a>b\;,
\eea
so that
\be\label{eq:prodgamma}
\textstyle\Gamma(b)=\Gamma(a)\prod\limits_{j}^{a:b}(j) \quad {\rm if } \quad a-b \in \mathds{Z}\;.
\ee
Furthermore, we define integration and differentiation operators~\cite{Moch:2001zr,Weinzierl:2004bn}
\bea
J^+(j,1)[f](x)&\equiv&\frac{1}{x^j}\int_0^x\d x' x'^{j-1}f(x') \; ,\nnb\\
\dps J^-(j,1)[f](x)&\equiv&\frac{1}{x^{j-1}}\frac{\d}{\d x}x^jf(x) \; ,\nnb
\eea
\bea
J^\pm (j,n)[f](x)\hs{-3}&\equiv&\hs{-3}\left[J^\pm (j,1)\right]\!\!\left[J^\pm (j,n-1)[f]\right]\!\!(x) \; ,\nnb\\
\eea
so that
\bea
\dps \frac{x^i}{(i+j)^n}&=& J^+(j,n)[y^i](x) \quad {\rm and}\nnb\\
\dps i^n x^i &=& J^-(0,n)[y^i](x) \; .
\eea
We now consider a HF of type $2_1^1$ and start from
\be
{}_2F_{1}(A_1, A_2;B_1;x)=1+ \nnb
\ee

\vs{-20}

\be\label{eq:2F1221}
\frac{\Gamma(B_1)}{\Gamma(A_1)\Gamma(A_{2})}\;\sum\limits_{i=1}^\infty\frac{\Gamma(A_1+i)\Gamma(A_2+i)}{\Gamma(B_1+i)\Gamma(i+1)}x^i
\ee
with
\be
A_1=a_1+\textstyle\frac{1}{2}+\alpha_1\epsilon,\quad A_2=a_2+\alpha_2 \epsilon,\nnb
\ee

\vs{-20}

\be
B_1=b_1+\textstyle\frac{1}{2}+\beta_1\epsilon \; .
\ee
We transform the $\Gamma$-functions in Eq.~(\ref{eq:2F1221}) by means of Eq.~(\ref{eq:prodgamma}) and arrive at
\be
\textstyle{}_2F_{1}(a_1+\frac{1}{2}+\alpha_1\epsilon, a_2+\alpha_2\epsilon;b_1+\frac{1}{2}+\beta_1\epsilon;x)
=1+\nnb
\ee

\vs{-20}

\be
\textstyle\frac{\Gamma(\frac{1}{2}+\beta_1\epsilon)}{\Gamma(\frac{1}{2}+\alpha_{1}\epsilon)\Gamma(1+\alpha_2\epsilon)}
\textstyle\frac{\prod\limits_{j}^{0:b_1}(j+\frac{1}{2}+\beta_1\epsilon)}
{\prod\limits_{j}^{0:a_1}(j+\frac{1}{2}+\alpha_1\epsilon)\prod\limits_{j}^{1:a_2}(j+\alpha_2\epsilon)} \nnb
\ee

\vs{-20}

\be
\textstyle\times\sum\limits_{i=1}^\infty
\underbrace{\textstyle\frac{\prod\limits_{j}^{0:a_1}(i+j+\frac{1}{2}+\alpha_1\epsilon)\prod\limits_{j}^{1:a_2}(i+j+\alpha_2\epsilon)}
{\prod\limits_{j}^{0:b_1}(i+j+\frac{1}{2}+\beta_1\epsilon)}}_{D} \nnb
\ee

\vs{-15}

\be
\hs{25}\times\frac{\Gamma(i+\frac{1}{2}+\alpha_1\epsilon)\Gamma(i+1+\alpha_2\epsilon)}
{\Gamma(i+\frac{1}{2}+\beta_1\epsilon)\Gamma(i+1)} \, x^i \; .
\ee
We now decompose of $D$ into partial fractions with respect to $\; i \;$ and obtain
\bea
D&=&\sum\limits_{j\ge0,n} \frac{C_{j,n}^+}{(i+j+\gamma\epsilon)^{n}}+\sum\limits_{j<0,n}
\frac{C_{j,n}^+}{(i+j+\gamma\epsilon)^{n}}\nnb\\
&&+\sum\limits_{j,n} \frac{C_{j,n}^{1/2}}{(i+\frac{1}{2}+j+\gamma\epsilon)^{n}}+\sum_n C_n^-i^n \, , \label{eq:Dpartial}
\eea
where $C_{j,n}^{+}$, $C_{n}^{-}$ and $C^{1/2}_{j,n}$ are polynomials in $\epsilon$.
In the first and third sum in Eq.~(\ref{eq:Dpartial}) we expand the denominator in $\epsilon$ and write the resulting expression in
terms of $J^+(j,n)$ and $J^+(j+\frac{1}{2},n)$.
In the last sum we express $i^n \, x^i$ in terms of $J^-(0,n)$. The second sum is conceptually also straightforward but results in quite
lengthy formulas which we omit here. We refer the reader to Ref.~\cite{Huber:2007dx} for details on this point.

The final formula reads
\be
{}_2F_{1}(a_1+\frac{1}{2}+\alpha_1\epsilon, a_2+\alpha_2\epsilon;b_1+\frac{1}{2}+\beta_1\epsilon;x)= \nnb
\ee

\vs{-18}

\be
1+\frac{\prod\limits_{j}^{0:b_1}(j+\frac{1}{2}+\beta_1\epsilon)}
{\prod\limits_{j}^{0:a_1}(j+\frac{1}{2}+\alpha_1\epsilon)\prod\limits_{j}^{1:a_2}(j+\alpha_2\epsilon)}\nnb
\ee
\be
\times\Big[\sum\limits_{j\ge0,n} \tilde C_{j,n}^+J^+(j,n)+\sum\limits_{j<0,n,\gamma} \tilde C_{j,n,\gamma}^+J^+(j,n,\gamma)\nnb
\ee
\be\label{eq:final211}
+\sum\limits_{j,n} \tilde C_{j,n}^{1/2}J^+(j+\frac{1}{2},n)+\sum_n C_n^-J^-(0,n)\Big]B \; ,
\ee
with the \textit{basis function} $B$ of this type,
\be
\dps
B={}_2F_1(\frac{1}{2}+\alpha_1\epsilon,1+\alpha_2\epsilon,\frac{1}{2}+\beta_1\epsilon,x)-1 \; .
\ee
Eq.~(\ref{eq:final211}) is most useful at the level of the expansion in $\epsilon$ since the various $\tilde C$ and
later on also $B$ enter this equation as expanded quantities. The generalization of this part of the algorithm to other types
$2^r_s$ and to ${}_PF_{P-1}$ with $P>2$ is straightforward. Moreover, this part of the algorithm is universal, i.\ e.\ type independent.
\subsection{Expansion of the basis function}
Here we make the ansatz
\bea
B&=&g(x)\bigg[1+\sum\limits_{j=1}^\infty \epsilon^j \\
&&\times \sum\limits_{\stackrel{s_1,...s_j=}{+,0,-}}c(s_1,...,s_j;x)H_{s_1,...,s_j}\left(f(x)\right)\bigg] \nnb
\eea
with $f(x)=\sqrt{x}$ for HFs of type $P^i_{i}$ and
\be
f(x)=i\sqrt{\frac{x}{1-x}} \quad {\rm or} \quad f(x)=\frac{1-\sqrt{1-x}}{1+\sqrt{1-x}}
\ee
for HFs of type $P^i_{i\pm 1}$~\cite{Jegerlehner:2002em,Davydychev:2003mv,Kalmykov:2006pu,Kalmykov:2006hu,Kalmykov:2007dk}. $H$ denotes
a harmonic polylogarithm (HPL)~\cite{Remiddi:1999ew}, and the weights ``$+$'' and ``$-$'' are, respectively, the sum and the difference
of the ordinary integer weights $\pm 1$~\cite{Maitre:2007kp}.
The function $g(x)$ is given by the value of the HF with the expansion parameter $\epsilon$ put to zero.
The coefficients $c(s_1,...,s_j;x)$ have the following properties. 
They are homogeneous of order $j$ in the $\alpha_i$, $\beta_i\,$; and symmetric in all $\alpha$ and $\beta$ parameters which
correspond to equal $a$ and $b$ parameters. Moreover, they must reduce to the coefficient of a reduced HF in the limit as one of the
$A_i$ becomes equal to one of the $B_k$. We also make an ansatz for the $x$-dependence of $c(s_1,...,s_j;x)$. This
ansatz depends on $f(x)$ and is rather simple. For type $2^1_1$ it is for instance a constant or $\sqrt{x}$ times a constant, depending
on whether we have an even or odd number of ``$+$'' weights in $\{s_1,\ldots,s_j\}$. 
We then insert the complete ansatz for $B$ into the differential equation for the HF ${}_PF_{P-1}$
\be
{\cal{D}} B =0 \; .
\ee
This yields after possible variable changes
\be
\sum\limits_{j=0}^\infty \epsilon^j\sum\limits_{l}\sum\limits_{\stackrel{s_1,...s_l=}{+,0,-}}{\cal
C}(s_1,...,s_l)H_{s_1,...,s_l}\left(y\right) \! = \! 0 \, .
\ee
The differential equation is satisfied if and only if all the coefficients ${\cal C}(s_i)$ vanish. The
coefficients $c(s_1,...,s_l;x)$ can be extracted from these conditions.
This part of the algorithm is a case-by-case approach, i.\ e.\ the expansion of the basis function must be derived separately for each
type $P^r_s$.
\subsection{Application of operators}
We are now left with the task of carrying out explicitly the integration and differentiation operations which now act on the expanded
basis function. Since the HPLs which occur in the expansion of the latter are iterated integrations over rational functions
they are well-suited for carrying out all required operations. The difficulties are to ensure 
the cancellation of all divergences $\propto 1/x^k$ at the lower integration limit, the integration of structures like
$1/\sqrt{x} \cdot H[\{...\},f(x)]$, as well as the need for introducing two new weights $w_1(t)=1/\sqrt{1-t^2}$ and
$w_2(t)=1/(t\sqrt{1-t^2})$ whose contributions cancel in the end.
\section{EXAMPLES}\label{sec:examples}
We implemented the above algorithm in the Mathematica package {\tt HypExp}~\cite{Huber:2005yg,Huber:2007dx}. The package\footnote{The package is
publicly available at \\ http://www-theorie.physik.uzh.ch/$\sim$maitreda/HypExp/} allows to expand arbitrary HFs
${}_PF_{P-1}(\{A_i\};\{B_j\};x)$ about integer parameters to arbitrary order in $\epsilon$, both for general argument $x$ and for unit
argument. The extension to half-integer parameters allows the expansion of HFs of types
\be
2^2_1,\;\; 2^1_1,\;\; 2^1_0,\;\; 2^0_1, \;\;
3^3_2,\;\; 3^2_2,\;\; 3^1_1,\;\; 3^1_0,\;\; 3^0_1, \;\;
4_1^1,\;\; 4_3^3 \nnb
\ee
also to arbitrary power in $\epsilon$, again for both general argument $x$ and $x=1$. In the following we give examples of multi-loop
diagrams which can be expanded by means of the package.
\subsection{Two-loop massive self-energy}
\begin{figure}[t]
\scalebox{.8}{\begin{picture}(340,120)
\Line(17,60)(137,60)
\LongArrow(17,60)(27,60)
\Text(18,70)[]{$p_m^2=m^2$}
\CArc(77,60)(40,0,360)
\CArc(77,60)(37.5,0,360)
\Text(77,90)[]{$M$}
\Text(77,30)[]{$M$}
\Text(77,54)[]{$m$}
\Line(155,75)(255,75)
\Line(155,77)(255,77)
\Vertex(205,76){1}
\Vertex(165,76){1}
\Vertex(245,76){1}
\CArc(185,65)(22.83,30,150)
\CArc(225,65)(22.83,30,150)
\CArc(205,86)(41.23,194.48,345.52)
\Text(185,94)[]{$n_2$}
\Text(225,94)[]{$n_2$}
\Text(188,69)[]{$n_1$}
\Text(222,69)[]{$n_1$}
\Text(205,38)[]{$n_3$}
\end{picture}}

\vspace*{-30pt}

\caption{Left panel: Two-loop massive self energy. Right panel: Three-loop on-shell HQET propagator diagram with mass.}
\label{fig:examplediagrams1}
\end{figure}
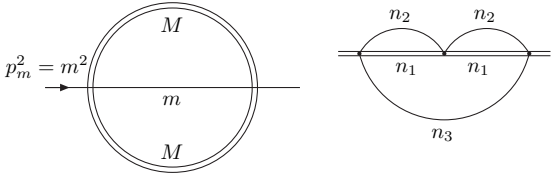

Our first example is the two-loop massive self-energy diagram depicted on the left in Fig.~\ref{fig:examplediagrams1}. It reads
\bea
I &=& \loopint{k_1} \loopint{k_2} \; \frac{1}{\lek k_2^2 -M^2 \rek} \\
&&\times \frac{1}{\lek (k_1-k_2)^2 -M^2 \rek \lek (k_1-p_m)^2 -m^2 \rek} \; ,\nnb
\eea
where an implicit $+i\eta$~($\eta>0$) is tacitly understood. This integral can be written in terms of HFs and assumes the very simple
form
\bea
I &=& - \ESGamma^2 \lek M^2 \rek^{1-2\eps} \, \frac{\Gamma^2(1-\eps)\Gamma^2(\eps)}{(1-\eps)} \nnb \\
&&\hspace*{-13pt}\times \Big\{ \frac{1}{1-2\eps} \, \pFq{3}{2}{\textstyle\frac{1}{2}\dps,1,-1+2\eps}{\textstyle\frac{1}{2}\dps
+\eps,2-\eps}{r} \nnb \\
&&+ r^{1-\eps} \pFq{3}{2}{1,\eps,\textstyle\frac{3}{2}\dps-\eps}{\textstyle\frac{3}{2}\dps,3-2\eps}{r} \Big\} \; ,
\eea
with $\ESGamma=1/(4\pi)^{D/2}/\Gamma(1-\eps)$ and $r=(m^2-i\eta)/M^2$. Expanding in $\eps$ we find
\be
I = - \ESGamma^2 \lek M^2 \rek^{1-2\eps} \, \Big\{ (1+\frac{r}{2}\dps) \frac{1}{\eps^2}\dps+(3+\frac{5}{4}\, r-r\ln r)\frac{1}{\eps}
\nnb
\ee

\vspace*{-20pt}

\be
 +\big(\frac{r}{2}\ln^2 r-\frac{(1-r)^2}{r} \, \mathrm{Li}_2(1-r)+(1-\frac{5}{2}\, r)\ln r \nnb
\ee

\vspace*{-20pt}

\be\label{eq:expandr}
\hspace*{5pt}+\frac{\pi^2}{3} \, r + \frac{\pi^2}{6 \, r} + 6 + \frac{11}{8} \, r \big) +{\cal{O}}(\eps) \Big\} \; ,
\ee
in agreement with Ref.~\cite{Argeri:2002wz}. For $m > M$ we find an expression for $I$ in terms of $\tilde r = (M^2-i\eta)/m^2$,
\bea
I &=& \ESGamma^2 \lek m^2 \rek^{1-2\eps} \, 2^{1-2\eps} \, \Gamma^2(1-\eps)\nnb\\
&&\hspace*{-33pt}\times \Big\{ \frac{\Gamma^2(\eps) \, \tilde r^{1-\eps} \, 2^{2\eps}}{(\eps-1) (1-2\eps)}
\pFq{3}{2}{\textstyle\frac{1}{2}\dps,1,2\eps-1}{\textstyle\frac{1}{2}\dps
+\eps,2-\eps}{\tilde r} \nnb \\
&&\hspace*{-13pt}+\frac{2^{2\eps} \, \Gamma^2(\eps) \, \tilde r^{2-2\eps}}{4(\eps-1)} \, 
\pFq{3}{2}{1,\textstyle\frac{3}{2}\dps-\eps,\eps}{\textstyle\frac{3}{2}\dps,3-2\eps}{\tilde r} \nnb \\
&&\hspace*{-13pt}+\frac{\pi \, \Gamma(-\textstyle\frac{1}{2}\dps+\eps)\Gamma(-\textstyle\frac{3}{2}\dps+2\eps)
\, \tilde r^{\frac{3}{2}-2\eps}}{\Gamma(\eps)} \nnb\\
&&\times \pFq{2}{1}{1-\eps,-\textstyle\frac{1}{2}\dps+\eps}{\textstyle\frac{5}{2}\dps
-2\eps}{\tilde r} \nnb\\
&&\hspace*{-13pt}+\frac{2^{1-2\eps} \, \Gamma^2(1-\eps)\Gamma(\eps)\Gamma(\textstyle\frac{3}{2}\dps-2\eps)\Gamma(-1+2\eps)}
{\sqrt{\pi}\,\Gamma(3-3\eps)} \nnb\\
&&\times \pFq{2}{1}{-2+3\eps,-\textstyle\frac{1}{2}\dps+\eps}{-\textstyle\frac{1}{2}\dps+2\eps}{\tilde r} \Big\} \; .
\eea
Upon expansion in $\eps$ we find
\be
I = - \ESGamma^2 \lek m^2 \rek^{1-2\eps} \, \Big\{ (\frac{1}{2}+\tilde r) \frac{1}{\eps^2}\dps+(\frac{5}{4}+3\, \tilde r-2\tilde r\ln
\tilde r)\frac{1}{\eps}
\nnb
\ee

\vspace*{-20pt}

\be
 +\big(\frac{\tilde r}{2}\, (2+\tilde r)\ln^2 \tilde r+(1-\tilde r)^2 \, \mathrm{Li}_2(1-\tilde r)-7\tilde r\ln \tilde r \nnb
\ee

\vspace*{-20pt}

\be\label{eq:expandrtilde}
\hspace*{5pt}+\frac{\pi^2}{3} + \frac{\pi^2}{6} \, \tilde r^2 + 6 \tilde r + \frac{11}{8}\big) +{\cal{O}}(\eps) \Big\} \; .
\ee
This result can also be obtained by analytic continuation of Eq.~(\ref{eq:expandr}) and proper inclusion of the analytic continuation
sign.
\subsection{Three-loop HQET master integral}
We elaborate only briefly on this integral since it has been discussed at length in Ref.~\cite{Grozin:2007ap}. It can be written as
\begin{eqnarray}
I_{n_1 n_2 n_3} &=& \frac{1}{i\pi^{d/2}}
\int \frac{I_{n_1 n_2}^2(p_0)\,d^d p}{(1-p^2-i0)^{n_3}}\,,
\nonumber\\
I_{n_1 n_2}(p_0) &=& \frac{1}{i\pi^{d/2}}
\int \frac{d^d k}{(-2(k_0+p_0)-i0)^{n_1}} \nnb\\
&& \times \frac{1}{(1-k^2-i0)^{n_2}}\,,
\label{I123}
\end{eqnarray}
and a closed form in terms of HFs with half-integer parameters was given in Ref.~\cite{Grozin:2007ap}. We focus here on a
particular combination of indices, namely
\be
\frac{I_{122}}{\Gamma^3(1+\epsilon)} =
 \frac{1}{2\epsilon^2} \Biggl[ {}
\frac{2 \Gamma^2(1-\epsilon) \Gamma^3(1+2\epsilon)}%
{\Gamma^2(1+\epsilon) \Gamma(1-2\epsilon) \Gamma(2+4\epsilon)}\nnb
\ee
\be
\hs{70}\times \, {}_3\!F_2 \left( \left.
\begin{array}{c}
\frac{1}{2}, 1+2\epsilon, -\epsilon\\
\frac{3}{2}+2\epsilon, 1-\epsilon
\end{array}
\right| 1 \right)
\nonumber
\ee
\be
-\frac{1}{1+2\epsilon} \;
{}_4\!F_3 \left( \left.
\begin{array}{c}
1, \frac{1}{2}-\epsilon, 1+\epsilon, -2\epsilon\\
\frac{3}{2}+\epsilon, 1-\epsilon, 1-2\epsilon
\end{array}
\right| 1 \right)
\nonumber
\ee
\be\label{I122}
 - 
\frac{\Gamma^2(1-\epsilon) \Gamma^4(1+2\epsilon)
\Gamma(1-2\epsilon) \Gamma^2(1+3\epsilon)}%
{\Gamma^4(1+\epsilon) \Gamma(1+4\epsilon)
\Gamma(1-4\epsilon) \Gamma(2+6\epsilon)}
\Biggr]\,.
\ee
In Ref.~\cite{Grozin:2007ap} we formulated the conjecture that the above expression is equal to
\begin{equation}
\frac{I_{122}}{\Gamma^3(1+\epsilon)} =
\frac{\pi^2}{3}
\frac{\Gamma^3(1+2\epsilon) \Gamma^2(1+3\epsilon)}%
{\Gamma^6(1+\epsilon) \Gamma(2+6\epsilon)}\,.
\label{dmfussballgott}
\end{equation}
The conjecture was formulated based on the agreement of the expansions up to the seventh order in $\eps$. Moreover, we performed further numerical checks at
the level of
the unexpanded expressions for various values of $\eps$ on the real axis and in the complex plane.
In Ref.~\cite{Grozin:2008mz} a reformulation of the above conjecture was given. Here we give another, alternative and very simple
reformulation of the conjecture.
\be
\MeijerG{33}{44}{1}{0,-\eps,\textstyle\frac{1}{2}\dps+\eps}{1+2\eps}{0,\eps,2\eps}{-\textstyle\frac{1}{2}\dps-\eps} \nnb
\ee
\be
= \frac{2^{2+4\eps} \, \pi^2 \, \Gamma(1-2\eps) \Gamma^2(1+2\eps)\Gamma^2(1+3\eps)}{3\, \Gamma(1-\eps) \Gamma^3(1+\eps)\Gamma(2+6\eps)}
 \; .\label{eq:reformulateconj}
\ee
Again, we have strong numerical evidence for this conjecture to hold true for any $\epsilon$ but we still lack an analytic proof of the
collaps of the MeijerG-function to mere $\Gamma$-functions.
\subsection{Three-loop master integral $A_{6,2}$}
Our third example is the three-loop master integral $A_{6,2}$ which is displayed on the left in
Fig.~\ref{fig:examplediagrams2}. It was calculated in Refs.~\cite{Huber:2007dx,Heinrich:2007at} using two different methods, one
based on HFs, the other one based on a two-dimensional Mellin-Barnes representation. Following the former method, we have
\begin{eqnarray}
\dps A_{6,2} &=& {\int \!\!\! \frac{d^D k}{\left(2\pi\right)^D}\!} {\int \!\!\! \frac{d^D l}{\left(2\pi\right)^D}\!} {\int \!\!\!
\frac{d^D r}{\left(2\pi\right)^D}\!} \; \frac{1}{\left( k+p_1\right)^2}\nnb \\
&\times& \hs{-5} \frac{1}{\left( k+l-p_2\right)^2  l^2 \, r^2 \left(r-k\right)^2 \,
\left( r-k-l\right)^2}\nnb\\
&=&\frac{{\cal N} \, 2^{8 \epsilon -2}\, \pi \,  \Gamma^2(1-3 \epsilon )
   \Gamma^5(1-\epsilon ) \Gamma (3 \epsilon )
  }{\epsilon  \, \Gamma (2-4 \epsilon ) \Gamma^2\!\left(\frac{3}{2}-2 \epsilon \right)}\nonumber
\eea
\be
\times\!\!\int\limits_0^1 \!\! ds \, s^{\epsilon -1} \, \bar s^{-3\epsilon }\left[\frac{s^{\eps}\,\Gamma (1-\epsilon )^2}{\Gamma (1-2
   \epsilon )}-{}_2F_1(\epsilon,-\epsilon ;1-\epsilon ;s)\right. \nnb
\ee
\be\label{eq:MIA62}
\times  _3F_2\!\left(\! 1-3
   \epsilon ,1-2 \epsilon ,1-\epsilon ;2-4 \epsilon
   ,\frac{3}{2}-2 \epsilon ;-\frac{\bar s^2}{4
   s}\right) ,
\ee
with $\bar s = 1-s$ and
\begin{equation}
{\cal N}=\frac{i \, (4\pi)^{3 \epsilon -6}}{ \Gamma^3(1-\epsilon )}\left(-q^2-i\eta\right)^{-3 \epsilon } \, .
\end{equation}
Eq.~(\ref{eq:MIA62}) can be expanded in $\eps$ at the level of the integrand, which yields HPLs of argument $-(1-s)/(1+s)$
that can be converted to HPLs of argument $s$ by applying (twice) the command {\tt
HPLConvertToSimplerArguments} from the {\tt HPL}~\cite{Maitre:2005uu,Maitre:2007kp} package. The next step is to expand the product of HPLs into a sum of
HPLs which we can then integrate by means of the integration routines of {\tt HPL}. This procedure
is not restricted to a specific depth of the expansion, so we could, in principle, expand $A_{6,2}$ to all orders. We expanded
$A_{6,2}$ up to transcendentality to eight,
\be
A_{6,2}=\frac{{\cal N}}{(1-5 \epsilon ) (1-4 \epsilon )
   \epsilon }\bigg[-2 \zeta_3-\epsilon\,\frac{7 \pi ^4}{180}\nnb
\ee
\be
   + \epsilon ^2\left(\frac{2}{3} \pi ^2 \zeta_3-10 \zeta_5\right)+\epsilon ^3\left(\frac{163 \pi^6}{7560}+76 \zeta_3^2\right)\nonumber
\ee
\be
 +\epsilon^4\left(\frac{55}{18} \pi ^4 \zeta_3+\frac{445
   \zeta_7}{2}\right)
   + \epsilon^5\left(-\frac{744}{5} \zeta_{5,3}-22 \pi ^2 \zeta_3^2\right.\nonumber
\ee
\be   
   \left.+1000 \zeta_3 \zeta_5+\frac{802183\pi ^8}{4536000}\right)+{\cal O}(\epsilon^6)\bigg]\, ,
\ee
where we have encountered a multiple zeta value in the last term.

\subsection{Four-loop tadpole with 3 massive lines}
Our last example is the four-loop tadpole with three equal massive lines displayed on the right in
Fig.~\ref{fig:examplediagrams2}. In Ref.~\cite{Gluza:2007rt}, a one-dimensional Mellin-Barnes
representation was derived for arbitrary powers of propagators.
We consider here the case of unit propagator powers and write
\be
T = \dps \int \! \left[dk_1\right] \! \int \! \left[dk_2\right] \! \int \! \left[dk_3\right]\! \int \!
  \left[dk_4\right] \frac{1}{\left[k_1^2-m^2\right]\left[k_3^2\right]}\nonumber
\ee

\vs{-15}

\be
\dps\times \frac{1}{\left[(k_1+k_2)^2-m^2\right]\left[(k_2+k_3+k_4)^2-m^2\right]\left[k_4^2\right]}\nonumber
\ee

\vs{-15}

\be
 \hspace*{10pt} = \dps -(m^2)^{3-4 \epsilon} \, e^{4\epsilon\gamma_E} \, \Gamma(1- \epsilon)/(1- \epsilon) \nonumber
\ee

\vs{-20}

\be
\times \MB{}{z} \, \Gamma^2(1-\epsilon-z) \Gamma(-z)\Gamma(2-2\epsilon-z)
 \nnb
\ee

\vs{-15}

\be
\times \frac{\Gamma(-2+3\epsilon+z)\Gamma(-3+4\epsilon+z)}{\Gamma(2-2\epsilon-2z)} ,
\label{eq:ex1mb}
\ee
with
$
\dps \left[dk\right]=\dps e^{\epsilon \gamma_E}/(i \pi^{D/2}) \, d^Dk
$
and $D=4-2\epsilon$ as usual.
Setting also the mass equal to unity and summing all residues of left poles of
$\Gamma$-functions in Eq.~(\ref{eq:ex1mb}), the result can be displayed in the following closed form
\be
T = \frac{2^{3-4\epsilon} \, e^{4\epsilon\gamma_E} \, \pi \, \Gamma^2(1-\epsilon)}{\sin(\pi \epsilon) \,
  \Gamma(2-\epsilon)}\nnb
\ee

\vs{-12}

\be
\times\left[\frac{\sqrt{\pi} \, \Gamma(\epsilon) \Gamma(-1+2\epsilon) \Gamma(-2+3\epsilon)}{\Gamma(2-\epsilon)
  \Gamma(-\frac{1}{2}+2\epsilon)}\right.\nonumber
\ee

\vs{-15}

\be
\hs{15}\times\pFq{3}{2}{\epsilon,2\epsilon-1,3\epsilon-2}{2-\epsilon,-\textstyle\frac{1}{2}\displaystyle+2\epsilon}{\textstyle
\frac{1}{4}\displaystyle}\nonumber
\ee

\vs{-20}

\be
-\frac{\Gamma(-\frac{1}{2}+\epsilon)\Gamma(-2+3\epsilon)\Gamma(-3+4\epsilon)}{\Gamma(-\frac{3}{2}+3\epsilon)}\nonumber
\ee

\vs{-15}

\be
\times
\pFq{3}{2}{2\epsilon-1,3\epsilon-2,4\epsilon-3}{\epsilon,-\textstyle\frac{3}{2}\displaystyle+3\epsilon}{\textstyle
\frac{1}{4}\displaystyle}\bigg] \, .
 \label{eq:ex1closed}
\ee
After some manipulations and simplifications on harmonic
polylogarithms, one gets for the expansion in $\epsilon$ up to the finite part
\be
T = \frac{1}{4\epsilon^4} + \frac{1}{\epsilon^3}+\left(\frac{97}{48}+\frac{\pi^2}{12}\right)\frac{1}{\epsilon^2}\nnb
\ee
\be
+ \left(\frac{833}{288}+\frac{\pi^2}{3}-\frac{\zeta_3}{3}\right)\frac{1}{\epsilon}+\frac{4177}{432}
+\frac{97\pi^2}{144}-\frac{4\zeta_3}{3}+\frac{\pi^4}{12}\nnb
\ee

\vs{-15}

\be
+\frac{1}{1728}\left[99+16\pi^2-24\,
\psi^{(1)}\!\!\left(\textstyle{\frac{1}{3}}\displaystyle\right)\right]^2+{\cal O}(\epsilon)\, ,
 \label{eq:ex1expanded}
\ee
in agreement with the findings of Refs.~\cite{Boughezal:2006xk,Gluza:2007rt,Faisst:2006sr}.
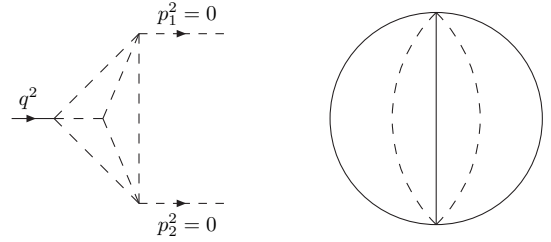
\begin{figure}[t]
\scalebox{.8}{\begin{picture}(340,120)
\Line(20,60)(30,60)
\LongArrow(10,60)(20,60)
\DashLine(30,60)(70,100){5}
\DashLine(30,60)(70,20){5}
\DashLine(70,20)(70,100){5}
\DashLine(30,60)(53,60){4}
\DashLine(53,60)(70,100){5}
\DashLine(53,60)(70,20){5}
\DashArrowLine(70,100)(110,100){5}
\DashArrowLine(70,20)(110,20){5}
\Text(93,110)[]{$p_1^2=0$}
\Text(93,10)[]{$p_2^2=0$}
\Text(18,70)[]{$q^2$}
\CArc(210,60)(50,0,360)
\Line(210,10)(210,110)
\DashCArc(260,60)(70.71,135,225){5}
\DashCArc(160,60)(70.71,315,45){5}
\end{picture}}

\vspace*{-20pt}

\caption{Left panel: Three-loop master integral $A_{6,2}$ with massless lines. Right panel: Four-loop tadpole diagram with two
massless lines (dashed) and three massive ones (solid) with equal masses.}\label{fig:examplediagrams2}
\end{figure}
\section{CONCLUSION AND OUTLOOK}\label{sec:conclusion}
The field of expanding hypergeometric functions ${}_PF_{P-1}$ about their parameters has acheived a quite sophisticated level. The
expansion about integer parameters involves ordinary HPLs only~\cite{Kalmykov:2007pf}, and algorithms for their all-order expansions
have been developed~\cite{Moch:2001zr,Weinzierl:2004bn,Kalmykov:2007pf} and implemented in computer algebra
systems~\cite{Weinzierl:2002hv,Huber:2005yg,Moch:2005uc}.

For the expansion about half-integral parameters there exist also several
algorithms~\cite{Weinzierl:2004bn,Kalmykov:2006pu,Kalmykov:2006hu,Kalmykov:2007dk,Huber:2007dx} for the all-order treatment of
many types $P^r_s$, and also implementations in various computer languages~\cite{Weinzierl:2002hv,Huber:2007dx}. However, not all types
$P^r_s$ can be expanded entirely in terms of ordinary HPLs, counterexamples involve for instance the types~$2^2_0$
and~$3^1_2$~\cite{Huber:2007dx}.

In the context of the expansion of HFs about other rational parameters some pioneering work was done by
Weinzierl~\cite{Weinzierl:2004bn} for the cases of so-called balanced fractions
\be
\hs{-2}\dps \frac{\Gamma(n+a_1-\frac{p_1}{q_1}+b_1 \epsilon)}{\Gamma(n+c_1-\frac{p_1}{q_1}+d_1 \epsilon)}  
\frac{\Gamma(n+a_2-\frac{p_2}{q_2}+b_2 \epsilon)}{\Gamma(n+c_2-\frac{p_2}{q_2}+d_2 \epsilon)}  ...
\ee
as well as single unbalanced rational numbers in numerator or denominator,
\be
\hs{-2}\dps \frac{\Gamma(n+1-\frac{p}{q}+b \, \epsilon)}{\Gamma(n+1+d \, \epsilon)} \quad \! {\rm or} \quad \!
\frac{\Gamma(n+a+b \, \epsilon)}{\Gamma(n+c-\frac{p}{q}+d \, \epsilon)}  \, .
\ee
However, up to now there are only few examples of HFs that contain other parameters than integral or half-integral ones. One important
application can be found in Ref.~\cite{Tarasov:2006nk}.

For the algorithm described here, the extension to arbitrary rational parameters is not a problem for the reduction part. However, the
expansion of the respective basis functions and the application of the differentiation and integration operators on the expanded basis
functions requires more conceptual work.
\section{ACKNOWLEDGMENTS}\label{sec:acknow}
I would like to thank the organizers of Loops~\&~Legs 2008 for creating a pleasant and inspiring atmosphere. Special thanks goes to
Daniel Ma{\^i}tre for a fruitful collaboration at all stages of the {\tt HypExp} project, and for a careful reading of the present
manuscript.


\begin{thebibliography}{9}
\bibitem{Moch:2001zr}
  S.~Moch, P.~Uwer and S.~Weinzierl,
  J.\ Math.\ Phys.\  {\bf 43}, 3363 (2002)
\bibitem{Weinzierl:2004bn}
  S.~Weinzierl,
  J.\ Math.\ Phys.\  {\bf 45} (2004) 2656
\bibitem{Kalmykov:2007pf}
  M.~Y.~Kalmykov, B.~F.~L.~Ward and S.~A.~Yost,
  JHEP {\bf 0711}, 009 (2007)
\bibitem{Weinzierl:2002hv}
  S.~Weinzierl,
  Comput.\ Phys.\ Commun.\  {\bf 145}, 357 (2002)
\bibitem{Bauer:2000cp}
  C.~W.~Bauer, A.~Frink and R.~Kreckel,
  arXiv:cs/0004015.
\bibitem{Huber:2005yg}
  T.~Huber and D.~Ma{\^i}tre,
  Comput.\ Phys.\ Commun.\  {\bf 175}, 122 (2006)
\bibitem{Moch:2005uc}
  S.~Moch and P.~Uwer,
  Comput.\ Phys.\ Commun.\  {\bf 174}, 759 (2006)
\bibitem{Davydychev:1992mt}
  A.~I.~Davydychev and J.~B.~Tausk,
  Nucl.\ Phys.\  B {\bf 397}, 123 (1993).
\bibitem{Broadhurst:1993mw}
  D.~J.~Broadhurst, J.~Fleischer and O.~V.~Tarasov,
  Z.\ Phys.\  C {\bf 60}, 287 (1993)
\bibitem{Tarasov:2006nk}
  O.~V.~Tarasov,
  Phys.\ Lett.\  B {\bf 638} (2006) 195
\bibitem{Davydychev:1998si}
  A.~I.~Davydychev and A.~G.~Grozin,
  Phys.\ Rev.\  D {\bf 59}, 054023 (1999)
\bibitem{Fleischer:2003rm}
  J.~Fleischer, F.~Jegerlehner and O.~V.~Tarasov,
  Nucl.\ Phys.\  B {\bf 672}, 303 (2003)
\bibitem{Jegerlehner:2002em}
  F.~Jegerlehner, M.~Y.~Kalmykov and O.~Veretin,
  Nucl.\ Phys.\  B {\bf 658}, 49 (2003)
\bibitem{Jegerlehner:2003py}
  F.~Jegerlehner and M.~Y.~Kalmykov,
  Nucl.\ Phys.\  B {\bf 676}, 365 (2004)
\bibitem{Davydychev:2000na}
  A.~I.~Davydychev and M.~Y.~Kalmykov,
  Nucl.\ Phys.\  B {\bf 605}, 266 (2001)
\bibitem{Davydychev:2003mv}
  A.~I.~Davydychev and M.~Y.~Kalmykov,
  Nucl.\ Phys.\  B {\bf 699}, 3 (2004)
\bibitem{Schroder:2005va}
  Y.~Schroder and A.~Vuorinen,
  JHEP {\bf 0506}, 051 (2005)
\bibitem{Bejdakic:2006vg}
  E.~Bejdakic and Y.~Schroder,
  Nucl.\ Phys.\ Proc.\ Suppl.\  {\bf 160}, 155 (2006)
\bibitem{Grozin:2007ap}
  A.~G.~Grozin, T.~Huber and D.~Ma{\^i}tre,
  JHEP {\bf 0707}, 033 (2007)
\bibitem{Argeri:2007up}
  M.~Argeri and P.~Mastrolia,
  Int.\ J.\ Mod.\ Phys.\  A {\bf 22}, 4375 (2007)
\bibitem{Kalmykov:2006pu}
  M.~Y.~Kalmykov,
  JHEP {\bf 0604}, 056 (2006)
\bibitem{Kalmykov:2006hu}
  M.~Y.~Kalmykov, B.~F.~L.~Ward and S.~Yost,
  JHEP {\bf 0702}, 040 (2007)
\bibitem{Kalmykov:2007dk}
  M.~Y.~Kalmykov, B.~F.~L.~Ward and S.~A.~Yost,
  JHEP {\bf 0710}, 048 (2007)
\bibitem{Huber:2007dx}
  T.~Huber and D.~Ma{\^i}tre,
  Comput.\ Phys.\ Commun.\  {\bf 178}, 755 (2008)
\bibitem{Vermaseren:1998uu}
  J.~A.~M.~Vermaseren,
  Int.\ J.\ Mod.\ Phys.\  A {\bf 14} (1999) 2037
\bibitem{GonzalezArroyo:1979df}
  A.~Gonzalez-Arroyo, C.~Lopez and F.~J.~Yndurain,
  Nucl.\ Phys.\  B {\bf 153} (1979) 161.
\bibitem{Blumlein:2003gb}
  J.~Blumlein,
  Comput.\ Phys.\ Commun.\  {\bf 159} (2004) 19
\bibitem{Remiddi:1999ew}
  E.~Remiddi and J.~A.~M.~Vermaseren,
  Int.\ J.\ Mod.\ Phys.\  A {\bf 15} (2000) 725
\bibitem{Maitre:2007kp}
  D.~Ma{\^i}tre,
  arXiv:hep-ph/0703052.
\bibitem{Argeri:2002wz}
  M.~Argeri, P.~Mastrolia and E.~Remiddi,
  Nucl.\ Phys.\  B {\bf 631} (2002) 388
\bibitem{Grozin:2008mz}
  A.~G.~Grozin,
  arXiv:0805.1474 [hep-ph].
\bibitem{Heinrich:2007at}
  G.~Heinrich, T.~Huber and D.~Ma{\^i}tre,
  Phys.\ Lett.\  B {\bf 662}, 344 (2008)
\bibitem{Maitre:2005uu}
  D.~Ma{\^i}tre,
  Comput.\ Phys.\ Commun.\  {\bf 174}, 222 (2006)
\bibitem{Gluza:2007rt}
  J.~Gluza, K.~Kajda and T.~Riemann,
  Comput.\ Phys.\ Commun.\  {\bf 177} (2007) 879
\bibitem{Boughezal:2006xk}
  R.~Boughezal and M.~Czakon,
  Nucl.\ Phys.\  B {\bf 755}, 221 (2006)
\bibitem{Faisst:2006sr}
  M.~Faisst, P.~Maierhoefer and C.~Sturm,
  Nucl.\ Phys.\  B {\bf 766}, 246 (2007)
\end{thebibliography}
\end{document}